\documentstyle[11pt,epsfig]{article}
\begin{document}

\title{Gravitational Waves: An Introduction}
\author{Indrajit Chakrabarty\footnote{\small Mehta Research Institute, Chhatnag Road, Jhusi. Allahabad. 211 019 INDIA}~\footnote{\scriptsize E-mail: indrajit@mri.ernet.in}}
\date{}
\maketitle

\begin{abstract}
In this article, I present an elementary introduction to the theory of
gravitational waves. This article is meant for students who 
have had an exposure to general relativity, but, results 
from general relativity used in the main discussion has been derived 
and discussed in the appendices. The weak gravitational field approximation 
is first considered and the linearized Einstein's equations are obtained. 
We discuss the plane wave solutions to these equations and consider
the transverse-traceless (TT) gauge. We then discuss the motion of
test particles in the presence of a gravitational wave and their
polarization. The method of Green's functions is applied to obtain 
the solutions to the linearized field equations in presence of
a nonrelativistic, isolated source. 

\end{abstract}
\vskip 1.5cm
\newpage

\section{Introduction}
In the past few years, research on the detection of gravitational waves has
assumed new directions. Efforts are now underway to detect gravitational 
radiation from astrophysical sources, thereby enabling researchers to
possess an additional tool to study the universe (See \cite{RMP} for 
a recent review). According to Newton's theory of gravitation, the binary
period of two point masses (e.g., two stars) moving in a bound orbit is
strictly a constant quantity. However, Einstein's general theory of relativity
 predicts that two stars revolving around each other in a 
bound orbit suffer accelerations, and, as a result, gravitational 
radiation is generated. Gravitational waves carry away energy and 
momentum at the expense of the orbital decay of two stars, thereby causing 
the stars to gradually spiral towards each other and giving rise to shorter 
and shorter periods. This anticipated decrease of the orbital period of a 
binary pulsar was first observed in PSR 1913+16 by Taylor and 
Weisberg (\cite{TW}). The observation supported the idea of gravitational 
radiation first propounded in 1916 by Einstein in the Proceedings
of the Royal Prussian Academy of Knowledge. Einstein showed that the
first order contributon to the gravitational radiation must be 
quadrupolar in a particular coordinate system. Two years later, he 
extended his theory to all coordinate systems. 
\par
The weak nature of gravitational radiation makes it very difficult to 
design a sensitive detector. Filtering out the noisy background to isolate 
the useful signal requires great technical expertise.
itself a field of research. Various gravitational wave detectors are 
fully/partially operational and we expect a certain result to 
appear from the observations in the near future.  
\par
This article gives an elementary introduction to the theory of gravitational
waves. Important topics in general relativity including a brief 
introduction to tensors and a derivation of Einstein's field equations
are discussed in the appendices. We first discuss the weak gravitational 
field approximation and obtain the linearized Einstein's field equations.
We then discuss the plane wave solutions to these equations in vacuum and the 
restriction on them due to the transverse-traceless (TT) gauge. The motion
of particles in the presence of gravitational waves and their polarization
is then discussed in brief. We conclude by applying the method of
Green's functions to show that gravitational radiation from matter at 
large distances is predominantly quadrupolar in nature. 
%
%
%
%
%
%
\newline
\section{The weak gravitational field approximation}
Einstein's theory of general relativity leads to Newtonian gravity in the
limit when the gravitational field is weak \& static and the particles in the
 gravitational field move slowly. We now consider a less 
restrictive situation where the gravitational field is weak but not static,
and there are no restrictions on the motion of particles in the 
gravitational field. In the absence of gravity, space-time is flat and
is characterised by the Minkowski metric, $\eta_{\mu\nu}$. A weak 
gravitational field can be considered as a small 'perturbation' on the
flat Minkowski metric\cite{Schutz},
\begin{equation}
g_{\mu\nu} = \eta_{\mu\nu} + h_{\mu\nu}, ~~~|h_{\mu\nu}| \ll 1
\end{equation}
Such coordinate systems are often called {\it Lorentz coordinate systems}. 
Indices of any tensor can be raised or lowered using $\eta^{\mu\nu}$ 
or $\eta_{\mu\nu}$ respectively as the corrections would be of higher 
order in the perturbation, $h_{\mu\nu}$. We can therefore write,
\begin{equation}
g^{\mu\nu} = \eta^{\mu\nu} - h^{\mu\nu}
\end{equation}
Under a background Lorentz transformation (\cite{Schutz}), the perturbation 
transforms as a second-rank tensor:
\begin{equation}
h_{\alpha\beta} = \Lambda_{\alpha}^{~~\mu} \Lambda_{\beta}^{~~\nu} ~h_{\mu\nu}
\end{equation}
The equations obeyed by the perturbation, $h_{\mu\nu}$, are obtained by
writing the Einstein's equations to first order. To the first order, the 
affine connection (See App. A) is,
\begin{equation}
\Gamma^{\lambda}_{~~\mu\nu} = \frac{1}{2} \eta^{\lambda\rho}[\partial_{\mu}
h_{\rho\nu} + \partial_{\nu}h_{\mu\rho} - \partial_{\rho}h_{\mu\nu}] + 
{\cal{O}}(h^2)
\end{equation} 
Therefore, the Riemann curvature tensor reduces to
\begin{equation}
R_{\mu\nu\rho\sigma} = \eta_{\mu\lambda}\partial_{\rho}\Gamma^{\lambda}_{~\nu\sigma} - \eta_{\mu\lambda}\partial_{\sigma}\Gamma^{\lambda}_{~\nu\rho}
\end{equation}
The Ricci tensor is obtained to the first order as
\begin{equation}
R_{\mu\nu} \approx R^{(1)}_{\mu\nu} = \frac{1}{2}\left[\partial_{\lambda}\partial_{\nu}h^{\lambda}_{~\mu} + \partial_{\lambda}\partial_{\mu}h^{\lambda}_{~nu} - \partial_{\mu}\partial_{\nu}h - \Box h_{\mu\nu}\right]
\end{equation}
where, $\Box \equiv \eta^{\lambda\rho}\partial_{\lambda}\partial_{\rho}$ is 
the D'Alembertian in flat space-time. Contracting again with $\eta^{\mu\nu}$,
the Ricci scalar is obtained as
\begin{equation}
R = \partial_{\lambda}\partial_{\mu} h^{\lambda\mu} - \Box h
\end{equation}
The Einstein tensor, $G_{\mu\nu}$, in the limit of weak gravitational field
is 
\begin{equation}
G_{\mu\nu} = R_{\mu\nu} - \frac{1}{2} \eta_{\mu\nu} R = \frac{1}{2}[\partial_{\lambda}\partial_{\nu}h^{\lambda}_{\mu} + \partial_{\lambda}\partial_{\mu}h^{\lambda}_{~\nu} - \eta_{\mu\nu}\partial_{\mu}\partial_{\nu}h^{\mu\nu} + \eta_{\mu\nu}\Box h - \Box h_{\mu\nu}]
\end{equation}
The linearised Einstein field equations are then
\begin{equation}
G_{\mu\nu} = 8 \pi G T^{\mu\nu}
\end{equation}
We can't expect the field equations (9) to have unique solutions as
any solution to these equations will not remain invariant under
a 'gauge' transformation. As a result, equations (9) will have infinitely 
many solutions. In other words, the decomposition (1) of $g_{\mu\nu}$ 
in the weak gravitational field approximation does not completely specify 
the coordinate system in space-time. When we have a system that is invariant
under a gauge transformation, we {\it fix} the gauge and work in a selected 
coordinate system. One such coordinate system is the {\it harmonic
coordinate system} (\cite{WB}). The gauge condition is
\begin{equation}
g^{\mu\nu}\Gamma^{\lambda}_{~~\mu\nu} = 0 
\end{equation}
In the weak field limit, this condition reduces to 
\begin{equation}
\partial_{\lambda} h^{\lambda}_{~~\mu} = \frac{1}{2}\partial_{\mu} h
\end{equation}
This condition is often called the {\it Lorentz gauge}. In this selected 
gauge, the linearized Einstein equations simplify to,
\begin{equation}
\Box h_{\mu\nu} - \frac{1}{2} \eta_{\mu\nu}\Box h = - 16 \pi G T^{\mu\nu}
\end{equation}
The `trace-reversed' perturbation, $\bar{h}_{\mu\nu}$, is defined as (\cite{Schutz}),
\begin{equation}
\bar{h}_{\mu\nu} = h_{\mu\nu} - \frac{1}{2}\eta_{\mu\nu} h
\end{equation}
The harmonic gauge condition further reduces to
\begin{equation}
\partial_{\mu} \bar{h}^{\mu}_{~\lambda} = 0
\end{equation}
The Einstein equations are then
\begin{equation}
\Box\bar{h}_{\mu\nu} = - 16 \pi G T^{\mu\nu}
\end{equation}
\vskip 1.0cm

\section{Plane-wave solutions and the transverse-traceless (TT) gauge}
From the field equations in the weak-field limit, eqns.(15), we 
obtain the linearised field equations in vacuum,
\begin{equation}
\Box \bar{h}_{\mu\nu} = 0
\end{equation}
The vacuum equations for $\bar{h}_{\mu\nu}$ are similar to the wave 
equations in electromagnetism. These equations admit the plane-wave 
solutions,
\begin{equation}
\bar{h}_{\mu\nu} = A_{\mu\nu} {\rm exp}(\iota k_{\alpha}x^{\alpha})
\end{equation}
where, $A_{\mu\nu}$ is a constant, symmetric, rank-2 tensor and $k_{\alpha}$
is a constant four-vector known as the {\it wave vector}. Plugging in the
solution (17) into the equation (16), we obtain the condition 
\begin{equation}
k_{\alpha}k^{\alpha} = 0
\end{equation}
This implies that equation (17) gives a solution to the wave equation (16)
if $k_{\alpha}$ is {\it null}; that is, tangent to the world line of a
photon. This shows that gravitational waves propagate at the speed of light.
The time-like component of the wave vector is often referred to as the 
{\it frequency} of the wave. The four-vector, 
$k_{\mu}$ is usually written as $k_{\mu} \equiv (\omega, {\bf k})$. 
Since $k_{\alpha}$ is null, it means that,
\begin{equation}
\omega^2 = |{\bf k}|^2
\end{equation}
This is often referred to as the {\it dispersion relation} for the 
gravitational wave. We can specify the plane wave with a number of 
independent parameters; 10 from the coefficients, $A_{\mu\nu}$ and three 
from the null vector, $k_{\mu}$. Using the harmonic gauge condition 
(14), we obtain,
\begin{equation}
k_{\alpha} A^{\alpha\beta} = 0 
\end{equation}
This imposes a restriction on $A^{\alpha\beta}$ : it is orthogonal 
({\it transverse}) to $k_{\alpha}$. The number of independent 
components of $A_{\mu\nu}$ is thus reduced to  six. We have to impose a 
gauge condition too as any coordinate transformation of the form
\begin{equation}
x^{\alpha^{\prime}} = x^{\alpha} + \xi^{\alpha}(x^{\beta})
\end{equation}
will leave the harmonic coordinate condition
\begin{equation}
\Box x^{\mu} = 0
\end{equation}
satisfied as long as
\begin{equation}
\Box \xi^{\alpha} = 0
\end{equation}
Let us choose a solution 
\begin{equation}
\xi_{\alpha} = C_{\alpha} {\rm exp}(\iota k_{\beta} x^{\beta})
\end{equation}
to the wave equation (23) for $\xi_{\alpha}$. $C_{\alpha}$ are constant 
coefficients. We claim that this remaining freedom allows us to convert from 
the old constants, $A^{({\rm old})}_{\mu\nu}$, to a set of new constants, 
$A^{({\rm new})}_{\mu\nu}$, such that
\begin{equation}
A^{({\rm new})~~\mu}_{\mu} = 0 ~~~~({\rm \it traceless})
\end{equation}
and
\begin{equation}
A_{\mu\nu} U^{\beta} = 0
\end{equation}
where, $U^{\beta}$ is some fixed four-velocity, that is, any constant 
time-like unit vector we wish to choose. The equations (20), (25) and 
(26) together are called the {\it transverse traceless} (TT) gauge 
conditions (\cite{Schutz}). Thus, we have used up all the available gauge
freedom and the remaining components of $A_{\mu\nu}$ must be physically 
important. The trace condition (25) implies that
\begin{equation}
\bar{h}^{TT}_{\alpha\beta} = h^{TT}_{\alpha\beta}
\end{equation}
\par
Let us now consider a background Lorentz transformation in
which the vector $U^{\alpha}$ is the time basis vector $U^{\alpha} = 
\delta^{\alpha}_{~0}$. Then eqn.(26) implies that $A_{\mu 0} = 0$ for all
$\mu$. Let us orient the coordinate axes so that the wave is travelling 
along the z-direction, $k^{\mu} \rightarrow (\omega, 0, 0, \omega)$. 
Then with eqn.(26), eqn.(20) implies that $A_{\alpha z} = 0$ for all
$\alpha$. Thus, $A^{TT}_{\alpha\beta}$ in matrix form is 
\begin{equation}
A^{TT}_{\alpha\beta} = \left( \begin{array}{cccc}
              0 & 0 & 0 & 0 \\
              0 & A_{xx} & A_{xy} & 0 \\
              0 & A_{xy} & -A_{xx} & 0 \\
              0 & 0 & 0 & 0 \end{array} \right)
\end{equation}
%
%
%
%
%
%
\section{Polarization of gravitational waves}
In this section, we consider the effect of gravitational waves on free 
particles. Consider some particles described by a single velocity field, 
$U^{\alpha}$ and a separation vector, $\zeta^{\alpha}$. Then, the separation 
vector obeys the geodesic equation (See App. A)
\begin{equation}
\frac{d^2\zeta^{\alpha}}{d\tau^2} = R^{\alpha}_{~\beta\gamma\delta} U^{\beta}
U^{\gamma}\zeta^{\delta}
\end{equation}
where, $U^{\nu} = {dx^{\nu}}/{d\tau}$ is the four-velocity of the two 
particles. We consider the lowest-order (flat-space) components of
$U^{\nu}$ only since any corrections to $U^{\nu}$ that depend on $h_{\mu\nu}$
will give rise to terms second order in the perturbation 
in the above equation. Therefore, $U^{\nu} = (1, 0, 0, 0)$ and 
initially $\zeta^{\nu} = (0, \epsilon, 0, 0)$.
Then to first order in $h_{\mu\nu}$, eqn. (29) reduces to 
\begin{equation}
\frac{d^2 \zeta^{\alpha}}{d \tau^2} = \frac{\partial^2 \zeta^{\alpha}}
{\partial t^2} = \epsilon R^{\alpha}_{~00x} = - \epsilon R^{\alpha}_{~0x0}
\end{equation}
Using the definition of the Riemann tensor, we can show that in the TT gauge,
\begin{eqnarray}
R^{x}_{~0x0} = & R_{x0x0} = & -\frac{1}{2} h^{TT}_{xx,00}\nonumber\\
R^{y}_{~0x0} = & R_{y0x0} = & -\frac{1}{2} h^{TT}_{xy,00}\nonumber\\
R^{y}_{~0y0} = & R_{y0y0} = & -\frac{1}{2} h^{TT}_{yy,00} = - R^{x}_{~0x0}
\end{eqnarray}
All other independent components vanish. This means that two particles 
initially separated in the x-direction have a separation vector which obeys 
the equation
\begin{equation}
\frac{\partial^2 \zeta^{x}}{\partial t^2} = \frac{1}{2} \epsilon \frac
{\partial^2}{\partial t^2}h^{TT}_{xx}, \frac{\partial^2\zeta^{y}}
{\partial t^2} = \frac{1}{2} \epsilon \frac{\partial^2}
{\partial t^2}h^{TT}_{xy}
\end{equation}
Similarly, two particles initially separated by $\epsilon$ in the y-direction
obey the equations
\begin{eqnarray}
\frac{\partial^2 \zeta^{y}}{\partial t^2}  = \frac{1}{2} \epsilon \frac{\partial^2}{\partial t^2}h^{TT}_{yy} =  -\frac{1}{2}\epsilon \frac{\partial^2}{\partial t^2}h^{TT}_{xx} \nonumber\\
\frac{\partial^2\zeta^{x}}{\partial t^2} =  \frac{1}{2} \epsilon 
\frac{\partial^2}{\partial t^2}h^{TT}_{xy}
\end{eqnarray}
We can now use these equations to describe the polarization of a gravitational 
wave. Let us consider a ring of particles initially at rest as in Fig. 1(a).
Suppose a wave with $h^{TT}_{xx}\neq 0, h^{TT}_{xy} = 0$ hits them. The 
particles respond to the wave as shown in Fig. 1(b). First the particles 
along the x-direction come towards each other and then move away from each 
other as $h^{TT}_{xx}$ reverses sign. This is often called $+$ polarization. 
If the wave had $h^{TT}_{xy}\neq 0$,
but, $h^{TT}_{xx} = h^{TT}_{yy} = 0$, then the particles respond as shown in 
Fig. 1(c). This is known as $\times$ polarization.
Since $h^{TT}_{xy}$ and $h^{TT}_{xx}$ are independent, the figures 
1(b) and 1(c) demonstrate the existence of two different states of 
polarisation. The two states of polarisation are oriented at an angle of
$45^{o}$ to each other unlike in electromagnetic waves were the two states of
polarization.
\vskip 1.0cm
\section{Generation of gravitational waves}
In section III, we obtained the plane wave solutions to the linearized 
Einstein's equations in vacuum, eqns.(16). To obtain the solution of
the linearised equations (15), we will use the Green's function method.
The Green's function, $G(x^{\mu} - y^{\mu})$, of the D'Alembertian 
operator $\Box$, is the solution of the wave equation in the presence of 
a delta function source:
\begin{equation}
\Box~G(x^{\mu} - y^{\mu}) = \delta^{(4)}(x^{\mu} - y^{\nu})
\end{equation}
where $\delta^{(4)}$ is the four-dimensional Dirac delta function. The 
general solution to the linearized Einstein's equations (15) can be 
written using the Green's function as
\begin{equation}
\bar{h}_{\mu\nu}(x^{\alpha}) = - 16\pi G\int d^4 y~G(x^{\alpha} - y^{\alpha})
T_{\mu\nu}(y^{\alpha})
\end{equation}
The solutions to the eqn.(34) are called {\it advanced} or {\it retarded}
according as they represent waves travelling backward or 
forward in time, respectively. We are interested in the retarded Green's 
function as it represents the net effect of signals from the past of the point
under consideration. It is given by
\begin{equation}
G(x^{\mu} - y^{\mu}) = - \frac{1}{4\pi |{\bf x} - {\bf y}|}
\delta\left[ |{\bf x} - {\bf y}| - (x^0 - y^0)\right]\times~\theta(x^0 - y^0)
\end{equation}
where, ${\bf x} = (x^1, x^2, x^3)$ and ${\bf y} = (y^1, y^2, y^3)$ and
$|{\bf x} - {\bf y}| = [\delta_{ij}(x^{i} - y^{i})(x^{j} - y^{j})]^{1/2}$.
$\theta(x^{0} - y^{0})`$ is the Heaviside unit step function, it equals
1 when $x^0 > y^0$, and equals 0 otherwise. Using the relation (36) in
(35), we can perform the integral over $y^0$ with the help of the delta 
function,
\begin{equation}
\bar{h}_{\mu\nu}(t, {\bf x}) = 4G\int d^3 y~\frac{1}{|{\bf x} - {\bf y}|}
T_{\mu\nu}(t - |{\bf x} - {\bf y}|, {\bf y}) 
\end{equation}
where $t= x^0$. The quantity
\begin{equation}
t_{\rm R} = t - |{\bf x} - {\bf y}|
\end{equation}
is called the {\it retarded time}. From the expression (37) for $\bar{h}_
{\mu\nu}$, we observe that the disturbance in the gravitational field at 
$(t, {\bf x})$ is a sum of the influences from the energy and momentum 
sources at the point $(t_{\rm R}, {\bf y})$ on the past light cone.
\par
Using the expression (37), we now consider the gravitational radiation
emitted by an isolated far away source consisting of very slowly moving 
particles (the spatial extent of the source is negligible compared 
to the distance between the source and the observer). The Fourier 
transform of the perturbation $\bar{h}_{\mu\nu}$ is 
\begin{equation}
{\tilde{\bar{h}}}_{\mu\nu} (\omega, {\bf x}) = \frac{1}{\sqrt{2\pi}}\int dt~{\rm exp}
(\iota\omega t)~\bar{h}_{\mu\nu}(t, {\bf x})
\end{equation}
Using the expression (37) for $\bar{h}_{\mu\nu} (t, {\bf x})$, we get
\begin{equation}
{\tilde{\bar{h}}}_{\mu\nu} = 4G\int d^3 y ~{\rm exp}(\iota\omega |{\bf x} - 
{\bf y}|)~\frac{\tilde{T}^{\mu\nu}(\omega, {\bf y})}{|{\bf x} - {\bf y}|}
\end{equation}
Under the assumption that the spatial extent of the source is negligible
compared to the distance between the source and the observer, we can replace
the term ${\rm exp}(\iota\omega |{\bf x} - {\bf y}|)/|{\bf x} - {\bf y}|$ 
in (40) by ${\rm exp}(\iota\omega{\rm R})/{\rm R}$. Therefore,
\begin{equation}
\tilde{\bar{h}}_{\mu\nu}(\omega, {\bf x}) = 4G~\frac{{\rm exp}(\iota\omega
{\rm R})}{{\rm R}}~\int d^3 y~\tilde{T}_{\mu\nu}(\omega, {\bf y})
\end{equation}
The harmonic gauge condition (14) in Fourier space is 
\begin{equation}
\partial_{\mu}\bar{h}^{~\mu\nu}(t, {\bf x}) = \partial_{\mu}\int d\omega~
{\tilde{\bar{h}}}^{~\mu\nu}~{\rm exp}(-\iota\omega t) = 0
\end{equation}
Separating out the space and time components,
\begin{equation}
\partial_0 \int d\omega~\tilde{\bar{h}}^{0\nu}(\omega, {\bf x}) {\rm exp}(-\iota\omega t) + \partial_{i}\int d\omega~\tilde{\bar{h}}^{i\nu}(\omega, {\bf x})
{\rm exp}(-\iota\omega t) = 0
\end{equation}
\vskip 2mm
\noindent
Or,
\begin{equation}
\iota\omega\tilde{\bar{h}}^{0\nu} = ~\partial_{i}\tilde{\bar{h}}^{i\nu}
\end{equation}
Thus, in eqn.(41), we need to consider the space-like components 
of ${\tilde{\bar{h}}}_{\mu\nu}(\omega, {\bf y})$. Consider,
\begin{eqnarray}
\int d^3 y~\partial_k \left(y^{i} \tilde{T}_{kj}\right) = \int d^3 y~\left( 
\partial_k y^i\right)\tilde{T}^{kj} + \int d^3 y~y^i\left(\partial_k\tilde{T}^{kj}\right)\nonumber
\end{eqnarray}
On using Gauss' theorem, we obtain, 
\begin{equation}
\int d^3 y~\tilde{T}^{ij}(\omega, {\bf y}) = - \int d^3 y~y^{i} \left(\partial_k \tilde{T}^{kj}\right)
\end{equation}	
Consider the Fourier space version of the conservation equation for 
$T^{\mu\nu}$, viz., $\partial_{\mu} T^{\mu\nu}(t, {\bf x}) = 0$. Separating
the time and space components of the Fourier transform equation, we have,
\begin{equation}
\partial_{i} \tilde{T}^{i\nu} = \iota\omega T^{0\nu}
\end{equation}
Therefore,
\begin{equation}
\int d^3 y~\tilde{T}^{ij}(\omega, {\bf y}) = \iota\omega\int d^3 y~y^{i}~ \tilde{T}^{0j} = \frac{\iota\omega}{2}\int~d^3 y~\left( y^{i}~\tilde{T}^{0j} + y^{j}~\tilde{T}^{0i}\right)
\end{equation}
Consider
\begin{equation}
\int~d^3 y~\partial_{l}\left(y_{i}~y_{j}\tilde{T}^{0l}\right) = \int~d^3 y\left[\left(\partial_l y^{i}\right)y^{j} + \left(\partial_l y^{j}\right)y^{i}\right]\tilde{T}^{0l} + \int d^3 y ~y^{i}~y^{j}\left(\partial_l \tilde{T}^{0l}\right)
\end{equation}
Simplifying the equation above, we obtain for the left hand side
\begin{eqnarray}
\int d^3 y~\left(y^{i}~\tilde{T}^{0j} + y^{j}\tilde{T}^{0i}\right) + 
\int d^3 y~y^{i}~y^{j}\left(\partial_l \tilde{T}^{0l}\right)\nonumber
\end{eqnarray}
Since the term on the left hand side of eqn.(47) is a surface term,
it vanishes and we obtain
\begin{equation}
\int d^3 y \left(y^{i}\tilde{T}^{0j} + y^{j}\tilde{T}^{0i}\right) = 
- \int d^3 y~y^{i}~y^{j}\left(\partial_l \tilde{T}^{0l}\right)
\end{equation}
Using the equations (46) and (48), we can write,
\begin{equation}
\int d^3 y~\tilde{T}^{ij}(\omega, {\bf y}) = \frac{\iota\omega}{2}\int
d^3 y~\partial_l\left(y^{i} y^{j}\tilde{T}^{0l}\right)
\end{equation}
Using the eqn(45), we can write,
\begin{equation}
\int d^3 y~\tilde{T}^{ij}(\omega, {\bf y}) = -\frac{\omega^2}{2}\int d^3 y~
y^{i}y^{j}~\tilde{T}^{00}
\end{equation}
We define the {\it quadrupole moment tensor} of the energy density of the 
source as
\begin{equation}
\tilde{q}_{ij}(\omega) = 3\int d^3 y~y^{i}y^{j}~\tilde{T}^{00}(\omega, {\bf y})
\end{equation}
In terms of the quadrupole moment tensor, we have
\begin{equation}
\int d^3 y~\tilde{T}^{ij}(\omega, {\bf y}) = - \frac{\omega^2}{6}~
\tilde{q}_{ij}(\omega)
\end{equation}
Therefore, the solution (41) becomes
\begin{equation}
\tilde{\bar{h}}_{ij}(\omega, {\bf x}) = 4G~\frac{{\rm exp}(\iota\omega{\rm R})}
{{\rm R}}\left(-~\frac{\omega^2}{6}\tilde{q}_{ij}(\omega)\right)
\end{equation}
Simplifying further,
\begin{equation}
\tilde{\bar{h}}_{ij}(\omega, {\bf x}) = -\frac{2}{3}~\frac{G\omega^2}{{\rm R}}
~{\rm exp}(\iota\omega{\rm R})~\tilde{q}_{ij}(\omega)
\end{equation}
Taking the Fourier transform of eqn.(54), and simplifying, we finally obtain
for the perturbation
\begin{equation}
\bar{h}_{ij}(t, {\bf x}) = \frac{2G}{3{\rm R}}~\frac{d}{dt^2}~q_{ij}(t_{{\rm R}})
\end{equation}
where, $t_{{\rm R}} = t - |{\bf x} - {\bf y}|$ is the retarded time. In the 
expression (54), we see that the gravitational wave produced by an isolated,
monochromatic and non-relativistic source is therefore proportional to the
second derivative of the quadrupole moment of the energy density at the 
point where the past light cone of the observer intersects the cone. 
The quadrupolar nature of the wave shows itself by the production of
shear in the particle distribution, and there is zero average translational
motion. The leading contribution to electromagnetic radiation comes from the 
changing dipole moment of the charge density. This remarkable 
difference in the nature of gravitational and electromagnetic radiation
arises from the fact that the centre of mass of an isolated system can't 
oscillate freely but the centre of charge of a charge distribution can. 
The quadrupole momnet of a system is generally smaller than the dipole 
moment and hence gravitational waves are weaker than electromagnetic 
waves. 
%
%
%
\section{Epilogue}
This lecture note on gravitational waves leaves several topics untouched.
There are a number of good references on gravitation 
where the inquisitive reader can find more about gravitational waves 
and their detection. I have freely drawn from various sources and I don't
claim any originality in this work. I hope I have been able to 
incite some interest in the reader about a topic on which there 
is a dearth of literature. 
\vskip 1cm
\centerline{\bf Acknowledgements}
\vskip 0.6cm
\noindent
This expository article grew out of a seminar presented at the end of 
the {\it Gravitation and Cosmology} course given by Prof. Ashoke Sen. 
I am grateful to all my colleagues who helped me during the 
preparation of the lecture. 
\newpage
\appendix
\vskip 0.5cm
\centerline{\large \bf Appendix A: Some topics in general theory of relativity}
\vskip 1.0cm
An event in relativity is characterised by a set of coordinates
$(t, x, y, z)$ in a definite coordinate system. Transformations between
the coordinates of an event observed in two different reference frames
are called {\it Lorentz transformations}. These transformations mix
up space and time and hence the coordinates are redefined so that all
of them have dimensions of length. We write $x^{0}\equiv ct, x^{1}\equiv x, x^{2}\equiv y, x^{3}\equiv z$ and a general component of a four
vector ($x^{0}, x^{1}, x^{2}, x^{3}$) as $x^{\mu}$. A Lorentz
transformation is then written as
\begin{equation}
x^{\mu} = \Lambda^{\mu}_{~~\nu} x^{\nu}
\end{equation}
where, 
\begin{equation}
\Lambda = \left( \begin{array}{cccc}
               \gamma & -\beta\gamma & 0 & 0\\
               -\beta\gamma & \gamma & 0 & 0\\
                0 & 0 & 1 & 0 \\
	        0 & 0 & 0 & 1\end{array}\right)
\end{equation}
At this point, it is useful to note the {\it Einstein summation convention}:
whenever an index appears as a subscript and as a superscript in an 
expression, we sum over all values taken by the index. Under a Lorentz
transformation, the spacetime interval $-(ct)^{2} + x^2 + y^2 + z^2$ 
remains invariant. The length of a four-vector is given by 
\begin{equation}
|{\bf x}| = -(x^{0})^2 + (x^{1})^2 + (x^{2})^2 + (x^{3})^2
\end{equation}
We never extract a square root of the expression (59) since $|{\bf x}|$
can be negative. Four-vectors that have negative length are called
{\it time-like}, while those with positive lengths are called 
{\it space-like}. Four-vectors with zero length are called {\it null}.
The notion of ``norm'' of a four-vector is introduced with the help of
the {\it Minkowski metric}:
\begin{equation}
\eta = \left( \begin{array}{cccc}
           -1 & 0 & 0 & 0 \\
	    0 & 1 & 0 & 0 \\
	    0 & 0 & 1 & 0 \\
            0 & 0 & 0 & 1 \end{array} \right)
\end{equation}
Then, we have,
\begin{equation}
|{\bf x}| = x^{\mu} \eta_{\mu\nu} x^{\nu}
\end{equation}
There are two kinds of vectors that are classified in the way they transform 
under a Lorentz transformation:
\begin{eqnarray}
{\rm Contravariant~~~:} x^{\mu} & = & \Lambda_{\nu}^{~~\mu} x^{\nu}\nonumber\\
{\rm Covariant~~~~~~~~~~:} x_{\mu} & = & \Lambda_{\mu}^{~~\nu} x_{\nu} 
\end{eqnarray}
Vectors are {\it tensors} of rank one. $\eta^{\mu\nu} (\eta_{\mu\nu})$ is
called the metric tensor; it is a tensor of rank two. There are other 
higher rank tensors which we will encounter later. If two coordinate 
systems are linked by a Lorentz transformation as:
\begin{equation}
x^{{\prime}~\nu} = \Lambda^{\nu}_{~~\mu} x^{\mu}
\end{equation}
then, multiplying both sides of the equation above by 
$\Lambda_{\nu}^{~~\kappa}$
and differentiating, we get,
\begin{equation}
\frac{\partial x^{\kappa}}{\partial x^{{\prime}~\nu}} = \Lambda_{\nu}^{~~\kappa} 
\end{equation}
Therefore, we see that
\begin{equation}
\frac{\partial}{\partial x^{{\prime}~\mu}} = \Lambda_{\mu}^{~~\nu}
\frac{\partial}{\partial x^{\nu}}
\end{equation}
Thus,
\begin{equation}
\partial_{\mu} \equiv \partial/\partial x^{\mu} = \left(
\frac{1}{c}\frac{\partial}{\partial t}, \frac{\partial}{\partial x}, 
\frac{\partial}{\partial y}, \frac{\partial}{\partial z}\right)
\end{equation}
transforms as a covariant vector. The differential operates on tensors to 
yield higher-rank tensors. A scalar $s$ can be constructed using the 
Minkowski metric and two four-vectors $u^{\mu}~{\rm and}~v^{\nu}$ as:
\begin{equation}
s =  \eta_{\mu\nu} u^{\mu}v^{\nu}
\end{equation}
A scalar is an invariant quantity under Lorentz transformations. 
Using the chain rule, 
\begin{equation}
d x^{{\prime~}\mu} = \frac{\partial x^{{\prime~}\mu}}{\partial x^{\nu}}
d x^{\nu}
\end{equation}
we have,
\begin{equation}
s = \left( \eta_{\mu\nu}\frac{\partial x^{\mu}}{\partial x^{{\prime}\kappa}}
\frac{\partial x^{\nu}}{\partial x^{{\prime}\lambda}}\right) 
u^{{\prime}\kappa} v^{{\prime}\lambda}
\end{equation}
If we define 
\begin{equation}
g_{\kappa\lambda} \equiv \eta_{\mu\nu}\frac{\partial x^{\mu}}{\partial x^{{\prime}\kappa}}\frac{\partial x^{\nu}}{\partial x^{{\prime}\lambda}}
\end{equation}
then,
\begin{equation}
s = g_{\kappa\lambda}~~ u^{{\prime}\kappa} v^{{\prime}\lambda}
\end{equation}
$g_{\kappa\lambda}$ is called the metric tensor; it is a symmetric, second-rank
tensor. 
\par To follow the motion of a freely falling particle, an {\it inertial}
coordinate system is sufficient. In an inertial frame, a particle at rest
will remain so if no forces act on it. There is a frame where particles 
move with a uniform velocity. This is the frame which falls freely in a
gravitational field. Since this frame accelerates at the same rate as the
free particles do, it follows that all such particles will maintain a uniform 
velocity relative to this frame. Uniform gravitational
fields are equivalent to frames that accelerate uniformly relative to 
inertial frames. This is the {\it Principle of Equivalence} between 
gravity and acceleration. The principle just stated is known as the
{\it Weak Equivalence Principle} because it only refers to gravity.
\par
In treating the motion of a particle in the presence of gravity, we
define the {\it Christoffel symbol} or the {\it affine connection}
as
\begin{equation}
\Gamma^{\mu}_{~~\alpha\beta} = \frac{1}{2} g^{\mu\nu}\left(
\frac{\partial g_{\nu\alpha}}{\partial x^{\beta}} +
\frac{\partial g_{\beta\nu}}{\partial x^{\alpha}} -
\frac{\partial g_{\alpha\beta}}{\partial x^{\nu}}\right)
\end{equation}
$\Gamma$ plays the same role for the gravitational field as the field 
strength tensor does for the electromagnetic field. Using the definition 
of affine connection, we can obtain the expression for the covariant 
derivative of a tensor:
\begin{equation}
{\cal{D}}_{\kappa} A^{\nu} \equiv \frac{\partial A^{\nu}}{\partial x^{\kappa}}
+ \Gamma^{\nu}_{~~\kappa\nu} A^{\alpha}
\end{equation}
It is straightforward to conclude that the covariant derivative of the metric 
tensor vanishes. The concept of ``parallel transport'' of a vector has 
important implications. We can't define globally parallel vector fields. We 
can define local parallelism. In the Euclidean space, a straight line is the
 only curve that parallel transports its own tangent vector. In curved space,
 we can draw ``nearly'' straight lines by demanding parallel transport of 
the tangent vector. These ``lines'' are called {\it geodesics}. A geodesic 
is a curve of extremal length between any two points. The equation of a 
geodesic is 
\begin{equation}
\frac{d^2 x^{\alpha}}{d \lambda^2} + \Gamma^{\alpha}_{~~\mu\beta}\frac{d x^{\mu}}{d\lambda}\frac{d x^{\beta}}{d\lambda} = 0
\end{equation}
The parameter $\lambda$ is called an {\it affine} parameter. A curve
having the same path as a geodesic but parametrised by a non-affine 
parameter is not a geodesic curve. The Riemannian curvature tensor is
defined as
\begin{equation}
R^{\mu}_{~\gamma\alpha\nu} = \frac{\partial \Gamma^{\mu}_{~\alpha\gamma}}{\partial x^{\nu}} - \frac{\partial \Gamma^{\mu}_{~\nu\gamma}}{\partial x^{\alpha}} + \Gamma^{\mu}_{~\nu\beta}~\Gamma^{\beta}_{~\alpha\gamma} - \Gamma^{\mu}_{~\alpha\beta}~\Gamma^{\beta}_{~\nu\gamma}
\end{equation} 
In a ``flat'' space,
\begin{equation}
R^{\mu}_{~\gamma\alpha\nu} = 0
\end{equation}
Geodesics in a flat space maintain their separation; those in curved spaces
don't. The equation obeyed by the separation vector $\zeta^{\alpha}$ in a 
vector field $V$ is
\begin{equation}
{\cal D}_{V}~{\cal D}_{V}~\zeta^{\alpha} = R^{\mu}_{~\gamma\alpha\nu} 
V^{\mu}~V^{\nu}~\zeta^{\beta}
\end{equation} 
If we differentiate the Riemannian curvature tensor and permute the indices,
we obtain the {\it Bianchi identity}:
\begin{equation}
\partial_{\lambda} R_{\alpha\beta\mu\nu} + \partial_{\nu} R_{\alpha\beta\lambda\mu} + \partial_{\mu} R_{\alpha\beta\nu\lambda} = 0
\end{equation}
Since in an inertial coordinate system the affine connection vanishes, 
the equation above is equivalent to one with partials replaced by covariant 
derivatives. The {\it Ricci tensor} is defined as
\begin{equation}
R_{\alpha\beta} \equiv  R^{\mu}_{~\alpha\mu\beta} = R_{\beta\alpha}
\end{equation}
It is a symmetric second rank tensor. The {\it Ricci scalar} (also known 
as {\it scalar curvature}) is obtained by a further contraction,
\begin{equation}
R \equiv R^{\beta}_{~\beta}
\end{equation}
The {\it stress-energy tensor} (also called {\it energy-momentum tensor}) 
is defined as the flux of the $\alpha$-momentum across a surface of
constant $x^{\beta}$. In component form, we have:
\begin{enumerate}
\item $T^{00}$ = Energy density = $\rho$
\item $T^{0i}$ = Energy flux (Energy may be transmitted by heat cinduction)
\item $T^{i0}$ = Momentum density (Even if the particles don't carry 
momentum, if heat is being conducted, then the energy will carry momentum)
\item $T^{ij}$ = Momentum flux (Also called {\it stress})
\end{enumerate}
%
%
%
%
%
%

\newpage
\centerline{\large \bf Appendix B: The Einstein field equation}
\vskip 1.0cm
The curvature of space-time is necessary and sufficient to describe gravity. 
The latter can be shown by considering the Newtonian limit of the geodesic 
equation. We require that
\begin{itemize}
\item the particles are moving slowly with respect to the speed of light,
\item the gravitational field is weak so that it may be considered as a 
perturbation of flat space, and,
\item the gravitational field is static.
\end{itemize}
In this limit, the geodesic equation changes to,
\begin{equation}
\frac{d^2 x^{\mu}}{d \tau^2}  +  \Gamma^{\mu}_{00}(\frac{dt}{d\tau})^2 = 0
\end{equation}
The affine connection also simplifies to
\begin{equation}
\Gamma^{\mu}_{00} = - \frac{1}{2} g^{\mu\lambda}\partial_{\lambda}g_{00}
\end{equation}
In the weak gravitational field limit, we can lower or raise the indices of
a tensor using the Minkowskian flat metric, e.g., 
\begin{equation}
\eta^{\mu\nu} h_{\mu\rho} = h^{\nu}_{~\rho}
\end{equation}
Then, the affine connection is written as 
\begin{equation}
\Gamma^{\mu}_{00} = - \frac{1}{2} \eta^{\mu\lambda}\partial_{\lambda}h_{00}
\end{equation}
The geodesic equation then reduces to
\begin{equation}
\frac{d^2 x^{\mu}}{d\tau^2} = \frac{1}{2}\eta^{\mu\lambda}\left(\frac{dt}{d\tau}\right)^2 \partial_{\lambda}h_{00}
\end{equation}
The space components of the above equation are
\begin{equation}
\frac{d^2 x^i}{d\tau^2} = \frac{1}{2}(\frac{dt}{d\tau})^2 \partial_i h_{00}
\end{equation}
Or,
\begin{equation}
\frac{d^2 x^i}{dt^2} = \frac{1}{2}\partial_i h_{00}
\end{equation}
The concept of an {\it inertial mass} arises from Newton's second law:
\begin{equation}
{\bf f} = m_i {\bf a}
\end{equation}
According to the the law of gravitation, the gravitational force exerted on
an abject is proportional to the gradient of a scalar field $\Phi$, known 
as the scalar gravitational potential. The constant of proportionality is the 
gravitational mass, $m_g$:
\begin{equation}
{\bf f}_g = - m_g {\bf \nabla}\Phi
\end{equation}
According to the Weak Equivalence Principle, the inertial and gravitational 
masses are the same,
\begin{equation}
m_i = m_g
\end{equation}
And, hence,
\begin{equation}
{\bf a} = - {\bf \nabla}\Phi
\end{equation}
Comparing equations (86) and (91), we find that they are the same if we 
identify,
\begin{equation}
h_{00} = - 2\Phi
\end{equation}
Thus,
\begin{equation}
g_{00} = - (1 + 2\Phi)
\end{equation}
The curvature of space-time is sufficient to describe gravity in the 
Newtonian 
limit as along as the metric takes the form (93). All the basic laws of 
Physics, beyond those governing freely-falling particles adapt to the 
curvature of space-time (that is, to the presence of gravity) when we are 
working in Riemannian normal coordinates. The tensorial form of any law
is coordinate-independent and hence, translating a law into the language
of tensors (that is, to replace the partial derivatives by the covariant 
derivatives), we will have an universal law which holds in all coordinate
systems. This procedure is sometimes called the Principle of Equivalence.
For example, the conservation equation for the energy-momentum tensor 
$T^{\mu\nu}$
in flat space-time, viz., 
\begin{equation}
\partial_{\mu} T^{\mu\nu} = 0
\end{equation}
is immediately adapted to the curved space-time as,
\begin{equation}
D_{\mu} T^{\mu\nu} = 0 
\end{equation}
This equation expresses the conservation of energy in the presence of a 
gravitational field. 
We can now introduce Einstein's field equations which governs how the metric 
responds to energy and momentum. We would like to derive an equation 
which will supercede the Poisson equation for the Newtonian potential:
\begin{equation}
\nabla^2 \Phi = - 4 \pi G \rho
\end{equation}
where, $\nabla^2 = \delta^{ij}\partial_{i}\partial_{j}$ is the Laplacian in 
space and $\rho$ is the mass density. A relativistic generalisation of this 
equation must have a tensorial form so that the law is valid in all coordinate
systems. The tensorial counterpart of the mass density is the 
energy-momentum tensor, $T^{\mu\nu}$. The gravitational potential should get 
replaced by the metric. Thus, we guess that our new equation will have 
$T^{\mu\nu}$ set proportional to some tensor which is second-order in the 
derivatives of the metric,
\begin{equation}
T^{\mu\nu} = \kappa A_{\mu\nu}
\end{equation}
where, $A_{\mu\nu}$ is the tensor to be found. The requirements on the 
equation above are:-
\begin{itemize}
\item By definition, the R.H.S must be a second-rank tensor. 
\item It must contain terms linear in the second derivatives or
quadratic in the first derivative of the metric.
\item The R.H.S must be symmetric in $\mu$ and $\nu$ as $T^{\mu\nu}$ is
symmetric.
\item Since $T^{\mu\nu}$ is conserved, the R.H.S must also be conserved.
\end{itemize}
The first two conditions require the right hand side to be of the form
\begin{equation}
\alpha R_{\mu\nu} + \beta R g_{\mu\nu} = T_{\mu\nu}
\end{equation}
where $R_{\mu\nu}$ is the Ricci tensor, $R$ is the scalar curvature and
$\alpha$ \& $\beta$ are constants. This choice is symmetric in $\mu$ and $\nu$
and hence satisfies the third condition. From the last condition, we 
obtain
\begin{equation}
g^{\nu\sigma}D_{\sigma}( \alpha R_{\mu\nu} + \beta R g_{\mu\nu}) = 0
\end{equation}
This equation can't be satisfied for arbitrary values of $\alpha$ and $\beta$.
This equation holds only if $\alpha/\beta$ is fixed. As a consequence of the 
Bianchi identity, viz.,
\begin{equation}
D^{\mu} R_{\mu\nu} = \frac{1}{2} D_{\nu} R
\end{equation}
we choose,
\begin{equation}
\beta = - \frac{1}{2}\alpha
\end{equation}
With this choice, the equation (42) becomes
\begin{equation}
\alpha(R_{\mu\nu} - \frac{1}{2} R g_{\mu\nu}) = T^{\mu\nu}
\end{equation}
In the weak field limit, 
\begin{equation}
g_{00} \approx -2\Phi
\end{equation}
the $00$-component of the equation(42), viz., 
\begin{equation}
-\alpha\nabla^2 g_{00} = T_{00}\Rightarrow 2\alpha\nabla^2\Phi = \rho
\end{equation}
Compare this result with Newtons equation (40), we obtain,
\begin{equation}
2a = \frac{1}{4 \pi G}
\end{equation}
Thus, we obtain the Einstein field equations in their final form as
\begin{equation}
R_{\mu\nu} - \frac{1}{2}R g_{\mu\nu} = 8 \pi G T^{\mu\nu}
\end{equation}

\newpage
\centerline{\large\bf Figures}
\vskip 1.5in

\begin{figure}[hbt]
\centerline{\epsfig{file=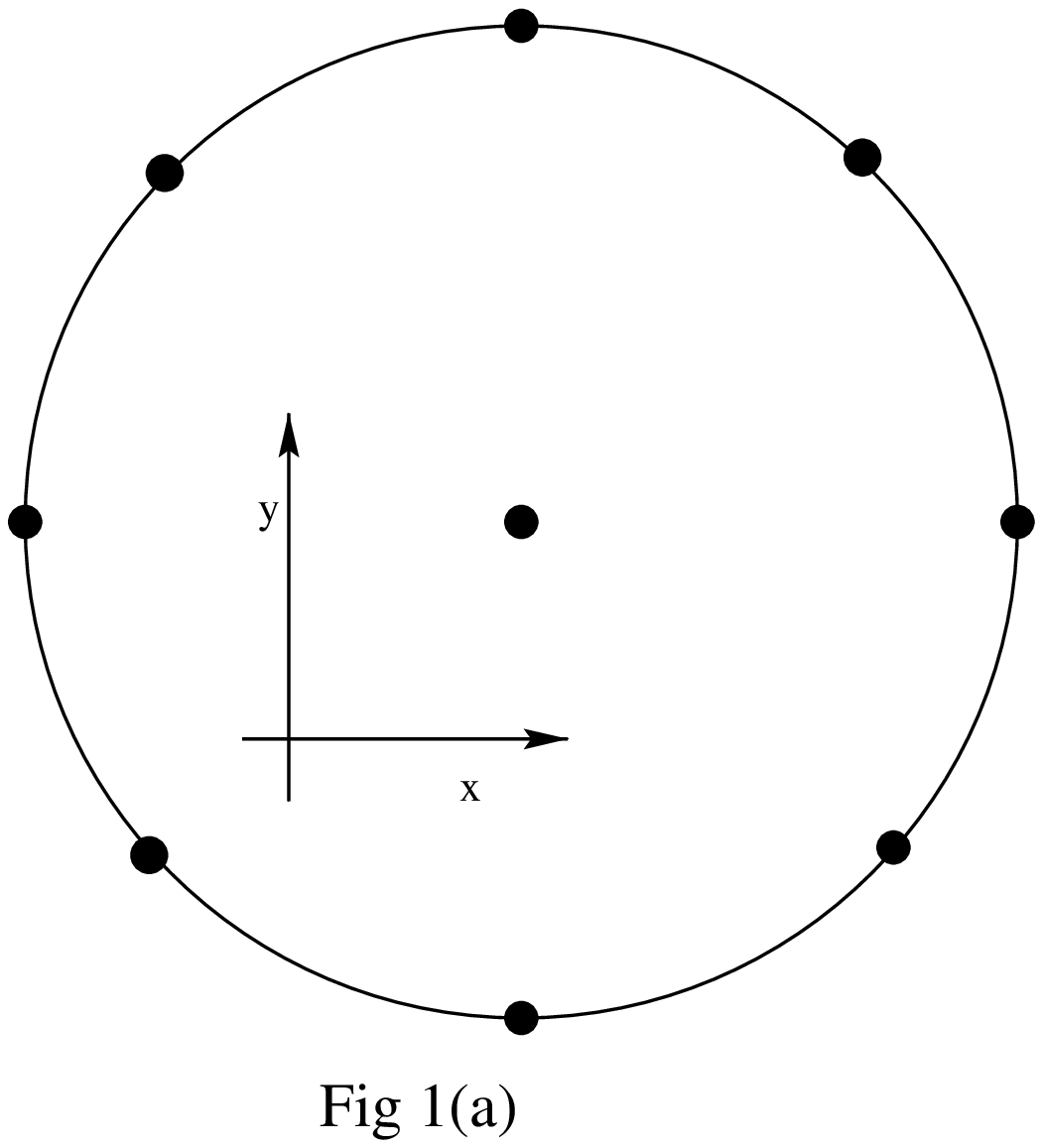,height=5cm}}
\vskip 2mm
\caption{
\scriptsize The initial configuration of test particles on a circle
before a gravitational wave hits them.}
\end{figure}

\begin{figure}[hbt]
\centerline{\epsfig{file=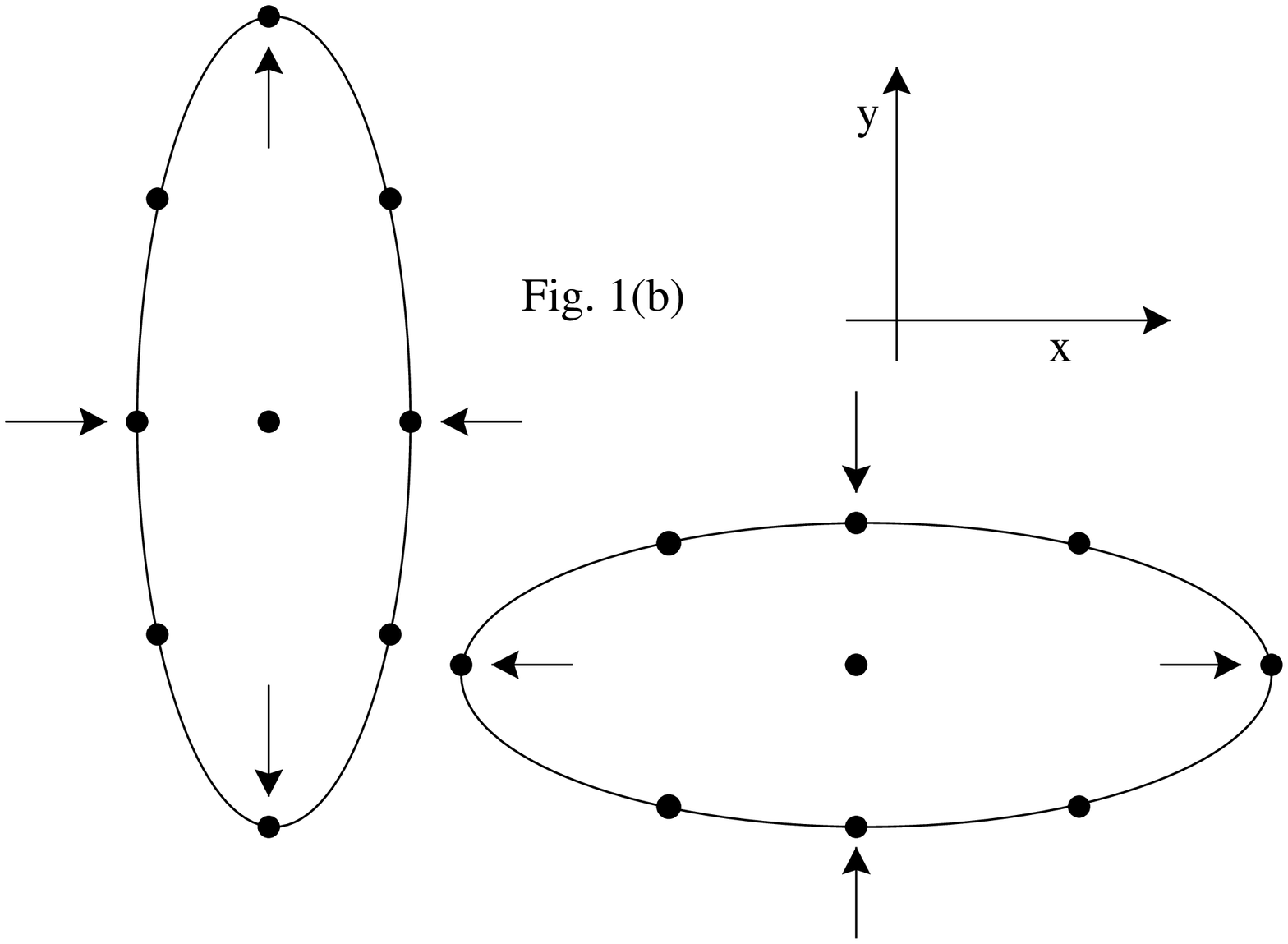,height=5cm}}
\caption{\scriptsize Displacement of test particles caused by the passage of a
gravitational wave with the + polarization. The two states are
separated by a phase difference of $\pi$.}
\end{figure}

\begin{figure}[hb]
\centerline{\epsfig{file=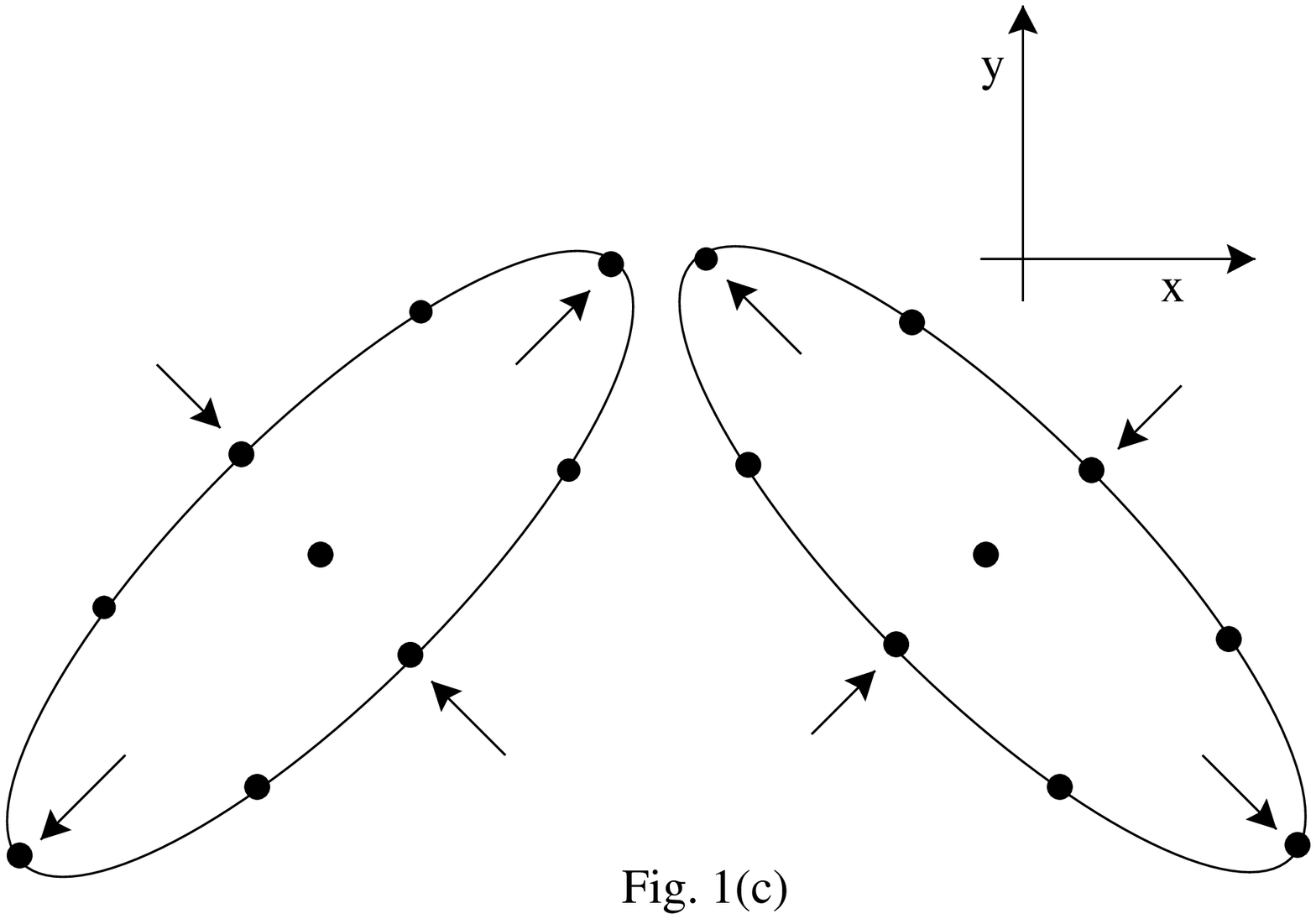,height=5cm}}
\caption{\scriptsize Displacement of test particles caused by the
passage of a gravitational wave with the $\times$ polarization.}
\end{figure}

\end{document}